\documentclass[a4paper]{ESASPCS13Style}
\usepackage{epsfig}

\begin{document}

\title{Mass-loss and Recent Spectral Changes in the Yellow Hypergiant $\rho$~Cassiopeiae}

\author{A. Lobel\inst{1}, J. Aufdenberg\inst{2}, I. Ilyin\inst{3}, \and
A. E. Rosenbush\inst{4}} \institute{Harvard-Smithsonian Center for Astrophysics, 60
  Garden Street, Cambridge, 02138 MA, USA  
\and National Optical Astronomy Observatory, P.O. Box 26732, Tucson, 85726 AZ, USA
\and Astrophysikalisches Institut Potsdam, An der Sternwarte 16, D-14482,
Potsdam, Germany
\and Main Astronomical Observatory, National Academy of Sciences of Ukraine,
Zabolotny str., 27, MSP Kyiv, 03680, Ukraine
}
\maketitle 

\begin{abstract}

The yellow hypergiant $\rho$ Cassiopeiae (F-G Ia0) has recently become
very active with a tremendous outburst event in the fall of 2000.
During the event the pulsating supergiant dimmed by more than a 
visual magnitude, while its effective temperature decreased from 7000 K to
below 4000 K over about 200 d, and we directly observed the largest
mass-loss rate of about 5\% of the solar mass in a single stellar
outburst so far. Over the past three years since the eruption
we observed a very prominent inverse P Cygni profile in Balmer
H$\alpha$, signaling a strong collapse of the upper atmosphere,
also observed before the 2000 event. Continuous spectroscopic
monitoring reveals that the H$\alpha$ line profile has transformed
into a P Cygni profile since June 2003, presently (Sept 2004)  
signaling supersonic expansion velocities up to $\sim$120 
$\rm km\,s^{-1}$ in the extended upper atmosphere, comparable to the 
2000 outburst. With the new fast atmospheric expansion many 
strong neutral atomic emission lines have appeared in the optical 
and near-IR spectrum over the past half year. 

Based on the very recent unique spectral 
evolution we observed the far-UV spectrum with the $FUSE$
satellite in July 2004. The $FUSE$ spectrum reveals that high-temperature 
plasma emission lines of O~{\sc vi} and C~{\sc iii} 
are absent in the hypergiant, also observed for the red supergiant 
$\alpha$ Ori (M2 Iab). On the other hand, we observe prominent
transition region emission lines in the smaller (less luminous)
classical Cepheid variable $\beta$ Dor (F-G Iab-Ia), indicating that
the mean atmospheric extension and surface gravity acceleration (as compared 
to effective temperature and atmospheric pulsation) play a major role 
for the formation of high-temperature stellar atmospheric plasmas. 
We present an overview of the recent spectral variability phases of $\rho$ Cas 
with enhanced mass-loss from this enigmatic cool star.

\end{abstract}

\section{Introduction}
\label{sec:int}

$\rho$ Cas is a cool {\em hypergiant}, one of the most luminous cool 
massive stars presently known. Yellow hypergiants are post-red supergiants, 
rapidly evolving toward the blue supergiant phase. They are among the prime 
candidates for progenitors of Type II supernovae in our Galaxy. This type of 
massive supergiant is very important to investigate the atmospheric dynamics 
of cool stars and their poorly understood wind acceleration mechanisms
that cause excessive mass-loss rates above $10^{-5}$ $\rm M_{\odot}\,yr^{-1}$ 
(Lobel et al. 1998). These wind driving mechanisms are also important to study the physical 
causes for the luminosity limit of evolved stars (Lobel 2001a; de Jager et
al. 2001). $\rho$ Cas is a rare bright cool hypergiant, which we are
continuously monitoring with high spectral resolution for more than a decade
(Lobel 2004). Its He-core burning phase is accompanied by tremendous episodic mass-loss
events, which we recently observed with a new outburst between 2000 July and 2001 April 
(Lobel et al. 2003a).

In quiescent pulsation phases $\rho$ Cas is a luminous late F-type supergiant
(Lobel et al. 1994).   
During a tremendous outburst of the star in 1945-47 strong absorption bands
of titanium-oxide (TiO) suddenly appeared in its optical spectrum, together with 
many low excitation energy lines, not previously observed.
These absorption lines, normally observed in M-type supergiants,
were strongly blue-shifted, signaling the ejection of a cool circumstellar 
gas shell. In July 2000 we observed the formation of new TiO bands during a strong 
$V$-brightness decrease by $\sim$$1^{\rm m}.4$ ({\it Fig. 1}). 
Our synthetic spectrum calculations show that $T_{\rm eff}$ decreased
by at least 3000 K, from 7250 K to $\simeq$4250~K, and the spectrum became comparable to the early
M-type supergiants $\mu$~Cep and Betelgeuse. The TiO bands  
reveal the formation of a cool circumstellar gas shell with $T$$<$4000
K due to supersonic expansion of the photosphere and upper atmosphere.
We observe a shell expansion velocity of $v_{\rm exp}$=35$\pm$2~$\rm
km\,s^{-1}$ from the TiO bands. 
From the synthetic spectrum fits to these bands we compute an exceptionally
large mass-loss rate of $\dot{M}$$\simeq$5.4\,$\times$\,$10^{-2}$~$\rm M_{\odot}\,yr^{-1}$,
comparable to the values estimated for the notorious outbursts of $\eta$
Carinae (Lobel et al. 2002, 2003b). 

\begin{figure*}[t]
  \begin{center}
    \epsfig{file=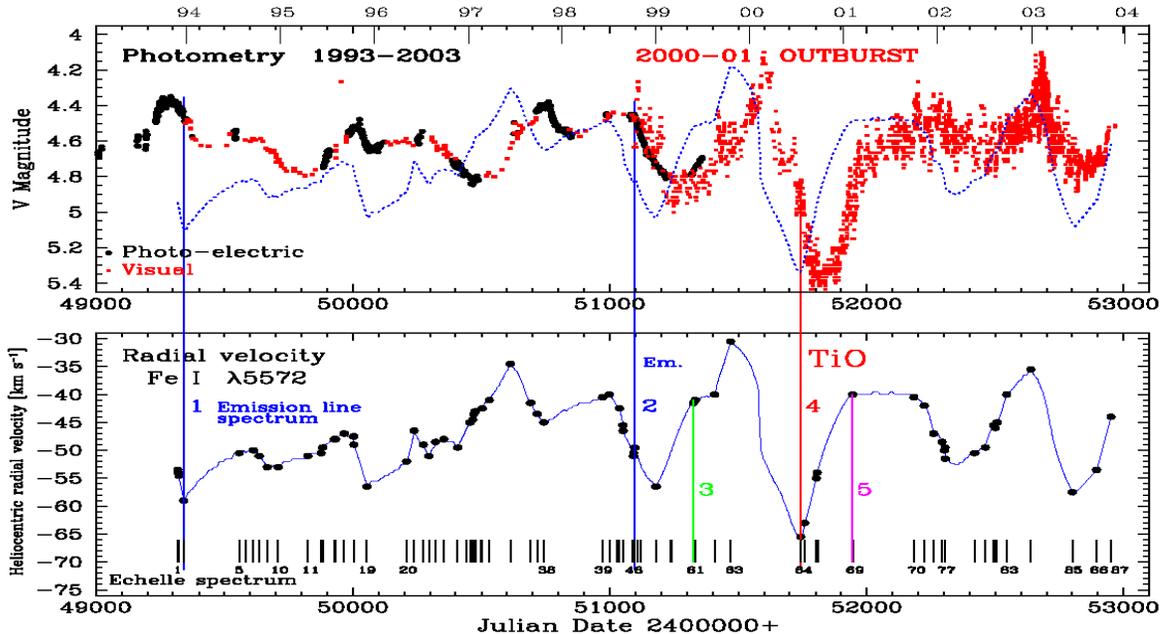, width=8cm}
  \end{center}
\caption{Visual brightness changes observed in $\rho$ Cas between 1993 and
  2003 (upper panel) are compared to radial velocity changes
(lower panel) observed from an unblended photospheric absorption 
line of Fe~{\sc i} $\lambda$5572. Fast atmospheric expansion in late 1999
and early 2000 precedes the deep visual outburst minimum of summer 2000.
Continuous high-resolution spectroscopic observations are marked with 
short vertical lines in the lower panel. A prominent emission line spectrum 
is observed during phases of fast atmospheric expansion in Dec. '93 and
Oct. '98 (labeled 1 and 2), and in Jan. '02 and July '04 (see Fig. 2).
\label{fig1}}
\end{figure*}

\section{Recent Continuous Spectroscopic Monitoring}
\label{sec:hal}

Over the past three years since the outburst we observed an unusually strong
inverse P Cygni profile in Balmer H$\alpha$, indicating a new collapse of $\rho$
Cas's extended upper atmosphere. Over the past year we observe how the 
H$\alpha$ line transformed from a prominent inverse P Cygni profile into a P Cygni
profile ({\it Fig. 3}). The P Cygni line shape of H$\alpha$ is very 
remarkable because it was not clearly observed during the 2000 brightness minimum. 
It indicates a new strong expansion of the upper atmosphere. We currently 
measure expansion velocities up to 120~$\rm km\,s^{-1}$ for the H$\alpha$
atmosphere. Our very recent high-resolution spectroscopic observations of April
2004 indicate that the P Cygni line shape further strengthens, resulting in a 
new and exceptional variability phase of enhanced mass-loss. 
This spectral evolution is expected since we observe that the 
radial velocity curve of the absorption portion of H$\alpha$ is at least twice 
longer than the (quasi-) period of photospheric absorption lines with
$P$$\sim$300-500 d (Lobel 2001b).  

Figure 5 shows dynamic spectra of H$\alpha$ and Fe~{\sc i} $\lambda$5572
observed between 1993 and late 2003. The white spots in H$\alpha$ are emission 
above the stellar continuum level. The radial velocity curves of the 
H$\alpha$ absorption core and the photospheric Fe~{\sc i} lines ({\it white dashed
lines}) reveal a velocity stratified dynamic atmosphere.
Notice the large blueshift of the Fe~{\sc i} lines during the outburst of mid
2000. The outburst is preceded by enhanced emission in the short
wavelength wing of H$\alpha$, while the Fe~{\sc i} line shifts far
redward. A strong collapse of the upper H$\alpha$ atmosphere and the lower
photosphere precedes the outburst event during the pre-outburst cycle of 1999 
(Lobel et al. 2003c) ({see \it Fig. 1}).

\begin{figure}[!bt]
  \begin{center}
    \epsfig{file=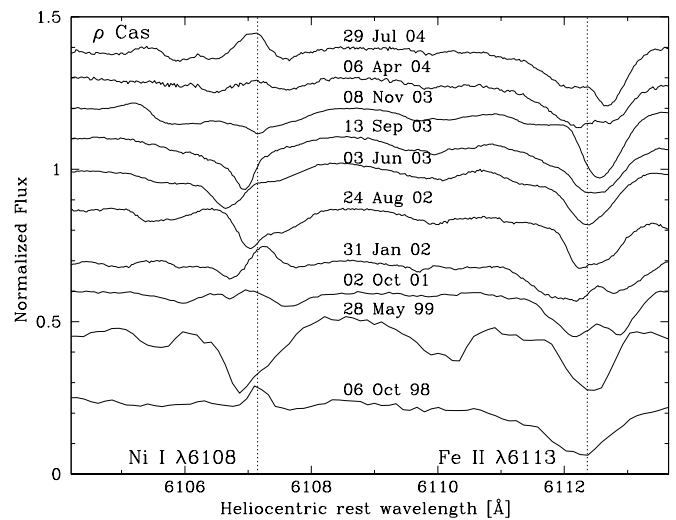, width=7cm}
  \end{center}
\caption{Spectral changes observed in $\rho$ Cas since mid 2004 reveal 
prominent atomic emission lines in the optical and near-IR spectrum.
The Ni~{\sc i} $\lambda$6108 line transforms from absorption 
into emission above the level of the stellar continuum during phases 
of fast atmospheric expansion in Fig. 1. Similar emission features
appear in the central core of photospheric absorption lines 
with small lower excitation energy as Fe~{\sc ii} $\lambda$6113, 
causing recurrent double (or split) absorption line cores. \label{fig2}}
\end{figure}

\begin{figure}[t]
  \begin{center}
    \epsfig{file=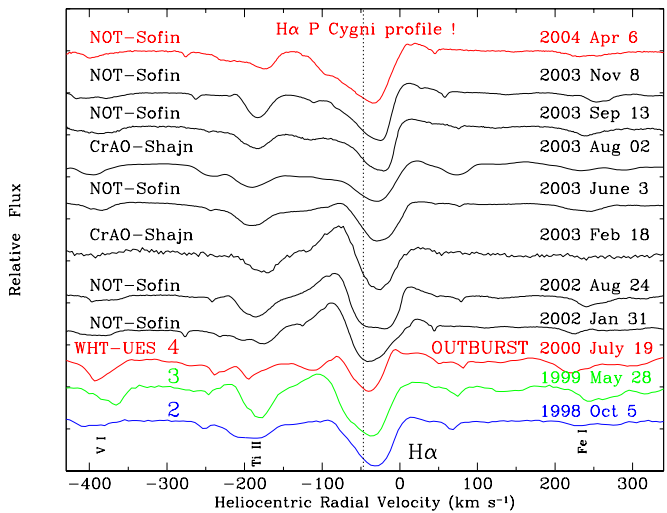, width=7cm}
  \end{center}
\caption{Line profile changes of H$\alpha$ in $\rho$ Cas reveal that 
following the outburst event of 2000 the upper atmosphere collapsed 
over a period of $\sim$2.5 years producing a strong inverse P Cygni 
profile. H$\alpha$ transformed into a P Cygni profile since mid 
2003, presently showing highly supersonic atmospheric expansion
comparable to the 2000 outburst. The vertical line is drawn at the 
stellar rest velocity of $-$47 $\rm km\,s^{-1}$.\label{fig3}}
\end{figure}
 
\begin{figure}[t]
  \begin{center}
    \epsfig{file=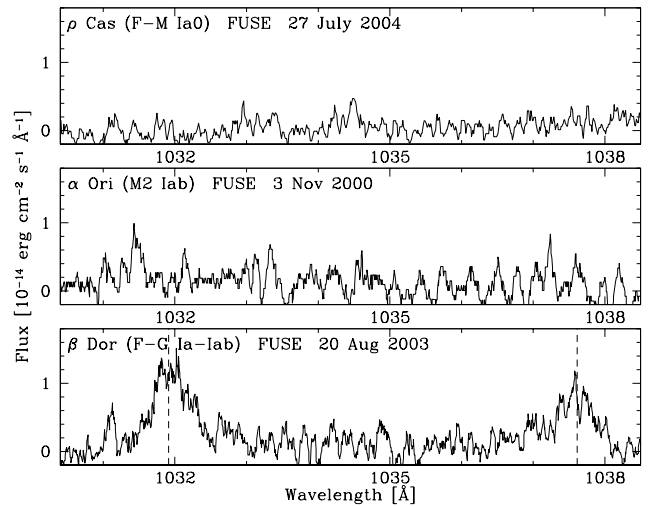, width=7cm}
  \end{center}
\caption{FUSE observations show that 
high-temperature emission lines of O~{\sc vi} 
$\lambda$1032 \& $\lambda$1037 are neither observed in yellow hypergiant 
$\rho$ Cas (upper panel), nor in red supergiant $\alpha$ Ori (middle panel). 
The lines are however clearly observed in the smaller supergiant Cepheid 
variable $\beta$ Dor (lower panel).\label{fig4}}
\end{figure}

\section{Far-UV, Optical, and Near-IR Emission Lines}
\label{sec:far}   

The optical and near-IR spectrum of $\rho$ Cas (Ia0e) only sporadically 
reveals prominent emission lines. The emission lines are due to neutral 
and singly ionized atoms.  
In a spectrum of $\rho$ Cas we observed on 29 July 2004, 
many exceptionally strong emission lines with low excitation potentials have 
re-appeared. The lines have only been observed in emission during previous phases 
of very fast atmospheric expansion in late '93 (Lobel 1997), fall 1998, 
and early 2002. Figure 2 shows how the Ni~{\sc i} $\lambda$6108 line transformed 
from absorption into emission with the fast expansion of the upper atmosphere
over the past half year. The line appears however in absorption during pulsation 
phases of photospheric collapse when $T_{\rm eff}$ increases. 
The formation of these rare emission lines therefore appears linked with an 
excitation phenomenon that can result from supersonic wind expansion 
during exceptional variability phases. A possible interpretation is 
that the emission spectrum is excited by a propagating circumstellar shock 
wave passing through the line formation region in the extended outer 
atmosphere (de Jager et al. 1997). 

The optical and near-IR emission line spectrum of $\rho$ Cas
is currently under investigation together with the 
far-UV spectrum observed with $FUSE$ on 27 July '04.
The $FUSE$ spectrum does not reveal the high temperature 
emission lines of O~{\sc vi} $\lambda$1032 \& $\lambda$1037 ({\em Fig 4}),
or the C~{\sc iii} $\lambda$977 line. Our far-UV observations 
also demonstrate that there is no evidence of a faint hot (e.g. white dwarf) 
companion star in $\rho$ Cas, also ruled out by our long-term
radial velocity and light curve monitoring. However, we clearly 
observe these warm transition region plasma emission lines in the classical 
Cepheid variable $\beta$ Dor (F-G Iab-Ia). Its smaller radius of
$R_{*}$=64~$\rm R_{\odot}$, compared to $\rho$ Cas ($R_{*}$$\simeq$400-500 $\rm
R_{\odot}$ with log($g$)=0.5-1.0), signals that the larger average atmospheric 
gravity acceleration is an important stellar parameter to sustain 
high-temperature plasmas in the outer atmosphere of less luminous
(pulsating) supergiants. This is supported by the observation that 
transition region plamsa emission lines are also absent in red 
supergiant $\alpha$ Ori with $R_{*}$=700 $\rm R_{\odot}$
and log($g$)=$-$0.5.       
 
\section{Conclusions}
\label{sec:con}

We present new recent spectral observations of the yellow hypergiant 
$\rho$ Cas that reveal enhanced dynamic activity of its pulsating 
atmosphere during the past three years after the outburst event 
of 2000. A strong inverse Balmer H$\alpha$ P Cygni profile that 
transformed into a P Cygni profile, currently signals supersonic wind expansion 
up to 120 $\rm km\,s^{-1}$ of the upper atmosphere with enhanced mass-loss. 
During the fast atmospheric expansion we observe the return of a 
prominent neutral and singly ionized atomic emission line spectrum in the
optical and near-IR. Very recent observations of the far-UV spectrum 
with $FUSE$ demonstrate the absence of high ion emission lines 
due to warm transition region plasma in the extended dynamic atmosphere of the hypergiant.   
    
\begin{figure*}[t]
  \begin{center}
    \epsfig{file=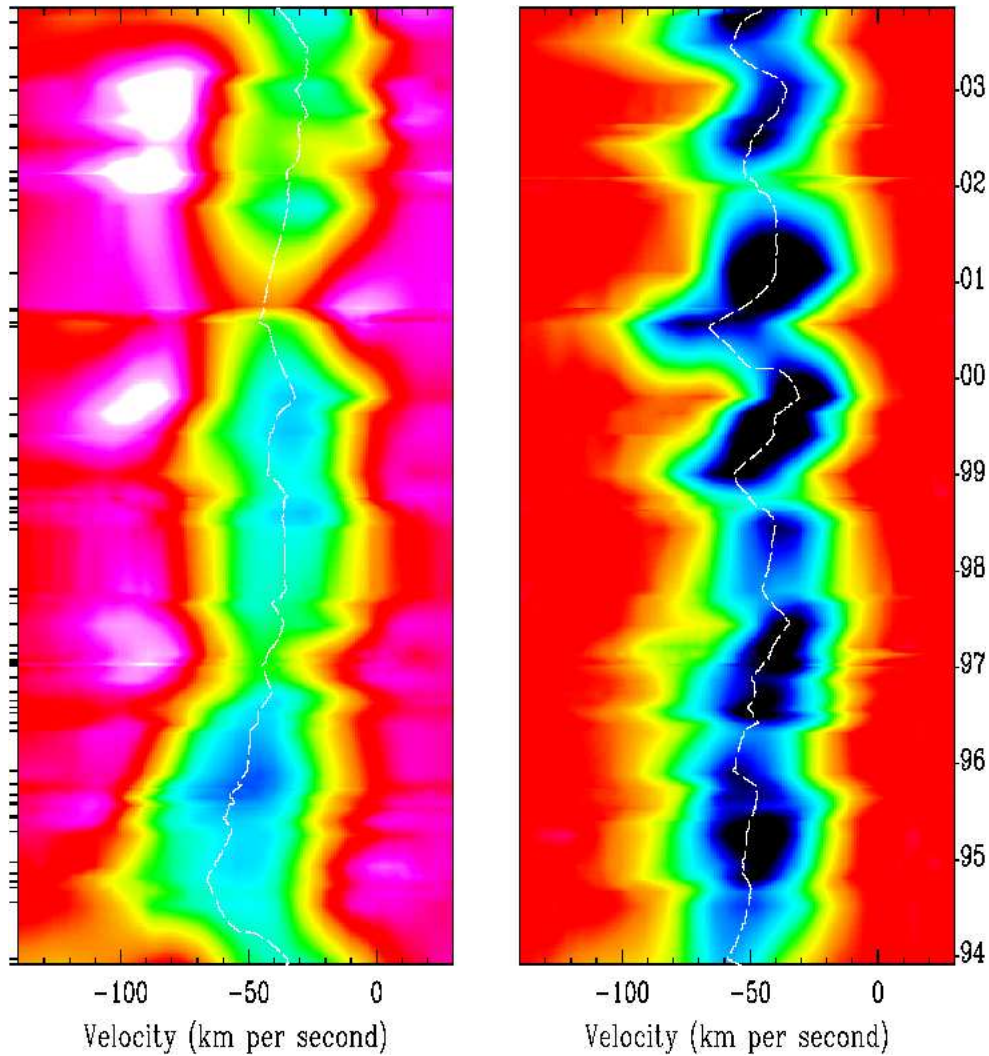, width=13cm}
  \end{center}
\caption{ Dynamic spectra of H$\alpha$ (panel left) and Fe~{\sc i}
$\lambda$5572 (panel right). The line profiles are linearly interpolated
between consecutive observation nights in the past decade, marked by the
left-hand tickmarks. Notice the strong blue-shift of the Fe~{\sc i} line
during the outburst of mid 2000. The outburst is preceded by very strong
emission (white spots) in the short-wavelength wing of H$\alpha$, while 
the absorption core extends longward, and the photospheric Fe~{\sc i} line
strongly red-shifts. A strong collapse of the upper and lower atmosphere 
precedes the outburst. Fast expansion of the upper H$\alpha$ atmosphere 
is also observed over the past year.\label{fig5}}
\end{figure*}  

\begin{acknowledgements}

This research is based in part on data obtained by the NASA-CNES-CSA $FUSE$ 
mission operated by the Johns Hopkins University. Financial support has 
been provided by NASA $FUSE$ grants GI-D107 and GI-E068.

\end{acknowledgements}

\end{document}